\documentclass{appolb}
\usepackage{epsfig}
\preprint{}

\begin{document}
\bibliographystyle{num}
\title{Size of the thermal source in relativistic heavy-ion collisions%
\thanks{Research supported in part by the Polish State Committee for
  Scientific Research, grant 2 P03B 059 25}%
}
\author{Piotr Bo\.zek
\address{The H. Niewodnicza\'nski Institute of Nuclear Physics, Polish Academy
of Sciences, PL-31342 Krak\'ow, Poland}
}
\maketitle
\begin{abstract}
The dependence of the size of the thermal source on the centrality 
in ultrarelativistic heavy-ion collisions is studied.
The interaction region consists of a well defined thermalized core, and of 
an outer mantle where the production scales with the number of participants.
The thermal source builds up in the region with the largest
 density of participants
in the transverse plane. 
Particle production in the thermalized core is enhanced in 
comparison to the wounded nucleon model.
The change of the  degree of strangeness saturation with centrality is
also discussed. We perform an estimate of high $p_\perp$  jet absorption  
finding that an increase of the absorption in the thermal core is compatible
with the data. 
\end{abstract}
  
\section{Introduction}
Relativistic heavy-ion collision experiments at RHIC energies
$\sqrt{s}=130$GeV
and $200$GeV reveal a large degree of thermalization in the produced system.
Particle spectra have thermal shapes, with a substantial amount of collective
flow. The presence of a strong transverse and elliptic collective flow
demonstrates the creation  of a strongly interacting phase in the history 
of the
system. This stage  can be described to a large extent by
 an ideal hydrodynamical 
evolution. Particle production in heavy-ion collisions is characterized by
the thermal parameters of the source~: its temperature, transverse flow and
possibly the degree of strangeness saturation. The temperature,
the chemical potentials  and the 
strangeness saturation at the chemical freeze-out can be extracted by fits
of a statistical hadronization model  to
the ratios of abundances of  particles produced in 
the central rapidity region. 
The description of particle spectra in transverse momentum 
requires the knowledge of the collective flow profile in the source and of
the temperature at the kinetic freeze-out.

Usually it is assumed that all the produced particles are
 emitted from  a single thermal source.
 One finds that  the parameters of the thermal
 freeze-out depend on the centrality. For the chemical freeze-out one observes
 a significant variation of the strangeness saturation factor in peripheral
 collisions \cite{xu,rafelski}. For the kinetic freeze-out a stronger
 collective flow and a
 lower freeze-out temperature comes out from the fits to the
 spectra of particles in  central collisions \cite{starspectra}.
A different approach allows for spatial variations of the parameters of the
 freeze-out surface surface, i.e. temperature and chemical potential gradients
\cite{budalund}.  

The extraction of the thermal parameters of the source and the description 
of its dynamics in 
hydrodynamical models are useful in the description of the evolution
of the dense matter produced in the collision. Such studies
 include  the modeling of the elliptic flow in non-central collisions. 
Quantitative results require the knowledge of the initial  conditions
in the simulation. Therefore the understanding of the variation of the size
and of the nature of the thermal source with centrality is crucial.

\section{Size of the thermal source}

Experimental data at the highest RHIC energies demonstrate the formation of a
thermalized system in $Au+Au$ collisions~: the spectra are thermal with a
collective transverse flow component, the elliptic flow is compatible 
with an ideal fluid dynamics, particle ratios can be described by universal
chemical freeze-out parameters. From the theoretical side we expect that at
high energy densities formed in the collision the system undergoes a phase
transition to the quark-gluon plasma. This phase transition could facilitate 
a complete  chemical equilibration. If the partonic plasma is strongly
interacting near the critical temperature, the system would behave as 
an ideal fluid. These feature should not vary with centrality if  the
thermalization and the evolution of the thermal source are dominated by the
phase transition. The kinetic freeze-out is given by the local density and
flow gradient and should not vary strongly with the impact parameter of the
collision either.

Therefore a reasonable assumption would be to have a thermal source with
the same thermal parameters at all centralities; the variation of the
particle production coming
 only from the change of the size of the source. 
It has been noticed that particle production per 
participant in heavy-ion collisions at RHIC energies is
significantly larger than in $p+p$ or $d+Au$ collisions 
\cite{Back:2004dy,Back:2004mr,Back:2002uc,Adler:2004zn}.
In the analysis of the experiments at RHIC energies
Glauber Monte Carlo calculations based on the wounded nucleon picture
permit to extract the number of participants (or
equivalently  wounded nucleons) used  to quantify 
the difference of the production per participant in $Au+Au$ or $d+Au$
 collisions
with respect to  $p+p$ interactions. 
Geometrical scaling predicts that the bulk of the particle production,
dominated by soft particles, scales with the number of wounded nucleons
\cite{abc}. It has been noticed that the density  of
charged particles at $\eta=0$ and the total number of charged 
particles produced
in $d+Au$ collisions 
scales with the number of participants~\cite{Back:2004mr}~:
\begin{equation}
\frac{dN_{dAu}}{d\eta}=N_{part}\frac{dN_{pp}}{d\eta} \ .
\end{equation}
Remarkably, this scaling is
 fulfilled in a more general way by the distributions of charged particles
in a broad range of  pseudorapidity \cite{bcscaling}.
On the other hand  particle production in $Au+Au$ collisions deviates from
 a simple superposition of independent production from all the wounded
 nucleons. The experimental data show a larger multiplicity  than predicted by
 the scaling of the  $p+p$ production by $N_{part}/2$. This and the 
 evidence of the thermalization discussed above point to the importance of
 the second stage of the collision. The main characteristic of this
 stage is the existence of a
 thermally equilibrated source.        Particle production
 from the thermal source occurs late  
and the production mechanism is far from the superposition of elementary
 collisions. In a system undergoing  Bjorken longitudinal expansion
the number of particles emitted at central rapidity is proportional to the 
transverse extension of the source. If the density of elementary collisions
 in the transverse plane is high one expect fast thermalization and eventually
 a phase transition to the quark-gluon plasma. In a geometrical Glauber model
 the decisive parameter is  the density of wounded nucleons in
the transverse plane. In peripheral collisions the density of colliding
 nucleons is small, and the production would be similar as in $d+Au$ collisions.
In semi-central and central collisions, at 
 the center of the interaction region
 the density of wounded nucleons is large and the transverse energy deposited
 is high enough to guaranty a fast thermalization. The boundary of the
thermal  source
 is given by the limiting density of wounded nucleons in the transverse plane
sufficient for the fast thermalization processes  to occur.
The observation of 
universal chemical freeze-out parameters  in $Au+Au$ collisions
and the indication of  the formation of a strongly interacting plasma
at near critical conditions, lead to the assumption that the boundary of the
 thermal source and of the quark-gluon plasma phase are the same. 
In the outer part of the interaction region the density of participants and
 therfore the density of the deposited energy is not
high enough to guaranty the 
approach of the critical point an the resulting fast thermalization. 
At RHIC
 energies the boundary in
 the interaction region between the thermalized
 source and the outer mantle is well defined by the  criterion of the phase
 transition. It turns out that the density of
participants in the outer mantle is low, they undergo at most $2.5$
binary  collisions per wounded nucleon pair. A similar number of 
binary collision is predicted for central $d+Au$ collisions. The number of 
 particles
 produced in $d+Au$ interactions (even in the most central collisions)
scales with the number of participants. The low number of binary collisions in
 the outer mantle   justifies the assumption that the number of charged
 particles produced in this way is proportional to the number of wounded
 nucleons in the mantle $N_{part}^{m}$.

The number of particles emitted
 from the source depends on the source size at the freeze-out. Since the 
freeze-out conditions are universal, the final number of particles produced
 should scale with the transverse energy deposited in the thermal source.
  The deposited energy is proportional to the 
number of wounded nucleons $N_{part}^{c}$ 
in this  high density core of the
 interaction region.
The density of charged particles produced in the central 
pseudorapidity region is
\begin{equation}
\frac{dN_{ch}}{d\eta}=\frac{N_{part}^{m}}{2}\frac{dN_{ch}^{pp}}
{d\eta}+
\frac{N_{part}^{c}}{2}\frac{dN_{ch}^{th}}{d\eta} \ ,
\end{equation}
where $\frac{dN_{ch}^{pp}}{d\eta}$ is the density of particles produced
produced in $p+p$ collisions and $\frac{dN_{ch}^{th}}{d\eta}$
is the density  per participant pair from the
 thermal source.
 The enhancement of 
the  particle production in the thermal source
can be effectively described  by one  parameter $\alpha$
\begin{equation}
\frac{dN_{ch}^{th}}{d\eta}=(1+\alpha)\frac{dN_{ch}^{pp}}{d\eta} \ .
\end{equation}

\begin{figure}
\centering
\includegraphics*[width=0.7\textwidth]{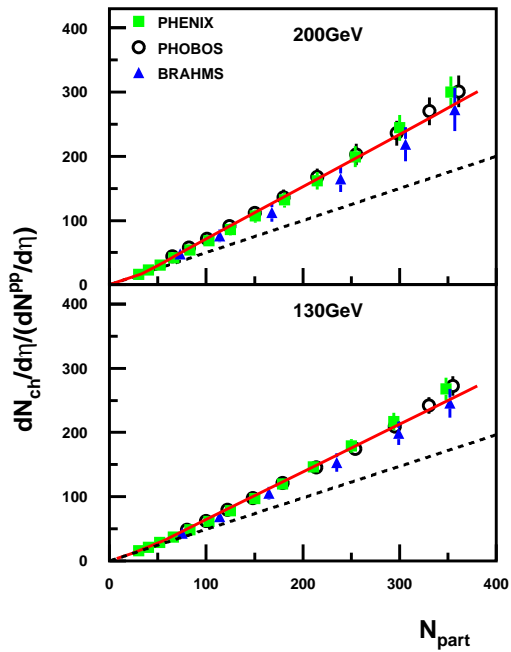}
\caption{Density of charged particles produced at central
 pseudorapidity in $Au+Au$
collisions scaled by the density of particles in $p+p$ collisions from
the BRAHMS \cite{Bearden:2001xw,Bearden:2001qq},
PHENIX \cite{Adler:2004zn} 
and  PHOBOS \cite{Back:2004dy,Back:2002uc} experiments 
 compared to the  predictions of the core-mantle model
 (solid line).
 The dashed line represents the  number of participant pairs.}
\label{specas}
\end{figure}

The number of wounded nucleons in the mantle is calculated from the formula
\begin{equation}
N_{part}^{m}=\int d^2s \frac{d\phi}{2\pi} \frac{d^2N_{part}}{ds^2}
\theta(d_{cut}-\frac{d^2N_{part}}{ds^2})\Theta(\phi) \ ,
\end{equation}
 where $\frac{d^2N_{part}}{ds^2}$ is the density of participants in the
 transverse plane, calculated from the
 Glauber Monte Carlo model. The integration
 is restricted to densities of participants lower than $d_{cut}$ and involves
 only particles emitted in directions other than the dense core (the function 
$\Theta(\phi)$ is one if the emission direction
 is not shadowed by the core and zero
 otherwise). The number of participants in the core is calculated
analogously
\begin{eqnarray}
N_{part}^{c}& =&\int d^2s \frac{d\phi}{2\pi} \frac{d^2N_{part}}{ds^2}
\theta(d_{cut}-\frac{d^2N_{part}}{ds^2})\left(1-\Theta(\phi)\right)
\nonumber \\
&+& \int d^2s \frac{d\phi}{2\pi} \frac{d^2N_{part}}{ds^2}
\theta(\frac{d^2N_{part}}{ds^2}-d_{cut}) \ ,
\end{eqnarray}
the main contribution comes from the region with the  density of participants
larger than $d_{cut}$. For the calculation of the density of wounded nucleons
we use Woods-Saxon nuclear density profiles
${1}/\left({1+\exp((r-R_{A})/a)}\right)$,
 with parameters $a=0.535$fm and $R_{A}=1.12A^{1/3}$fm.
The cross sections are $\sigma=42$mb and $41$mb for $\sqrt{s}=200$GeV and
$130$GeV respectively. 

The results of the calculations for the scaled particle pseudorapidity density
\begin{equation}
\frac{dN_{AuAu}}{d\eta}/\frac{dN_{pp}}{d\eta}
\end{equation}
 are presented in Figure \ref{specas}.
Published experimental data  from the BRAHMS, PHENIX and PHOBOS collaborations
are shown. The density of charged particles divided by the $p+p$ density 
increases faster than the number of participant pairs  
shown by the dashed line. Moreover this increase is not exactly linear in 
$N_{part}$. There is a strong increase of $dN/d\eta$ around $30-100$
participants and then a gradual increase of the 
slope up to the highest centralities. We can reproduce this effect in 
the core-mantle model with the parameters $d_{cut}=2$fm$^{-2}$ and 
$\alpha=0.65$ at $\sqrt{s}=200$GeV and $d_{cut}=2.2$fm$^{-2}$ and 
$\alpha=0.5$ at $\sqrt{s}=130$GeV. At centralities corresponding to
$N_{part}=40$ the thermalized core appears in the interaction region, 
causing a fast increase of the  scaled particle density.
For more central collisions the contribution of the 
thermal core in the  total
production increases gradually up to $95\%$ at zero impact parameter. This
leads to a slightly faster than linear increase of the  particle density
with the number of participants, similar as seen in the data. 
The increase of the production per participant pair in the core $1+\alpha$ and
the decrease of the cutoff density of participants $d_{cut}$ with energy
indicate that the wounded nucleons get effectively ``fatter'' in the the
transverse plane and deposit a larger transverse energy at higher
 collision energy. At SPS energies the mantle part of the interaction region
 is dominant. The number of binary collision in the mantle increases and the 
assumption of the $N_{part}^{m}$ scaling of the production in the mantle 
breaks down.

\section{Strangeness}

                     Following the arguments from the previous section 
one can argue that  the production of identified strange particles
in heavy-ion collisions at RHIC energies occurs by two different mechanisms 
in the core and in the mantle of the interaction region.
In studies with
a single thermal source different degree of strangeness equilibration
as function of centrality is seen 
\cite{xu,rafelski}.
In the core-mantle model we choose a different approach. Since the physics of
the thermal core is dominated by the phase transition we assume 
a complete equilibration of strangeness in the thermal source, this is
consistent with the data on particle ratios for the most central collisions.

\begin{figure}
\centering
\includegraphics*[width=0.7\textwidth]{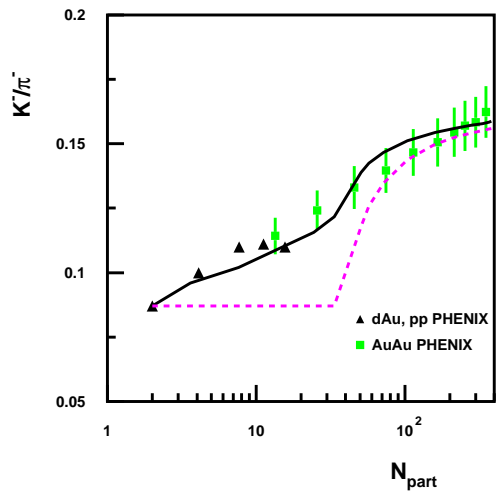}
\caption{$K^-/\pi^-$ ratio as a function of centrality. The 
data are from the PHENIX 
collaboration \cite{phenixstrange,phenixmata}.
The solid line denotes the sum of the emission from  the  thermal core
 and of a  contribution from $p+p$ collisions scaled with the 
number of collisions in the mantle (equation \ref{scaledau}).
 The dashed line represents the
results of a calculation with a thermal source and a scaled 
 $p+p$ contribution
for the  production in the mantle.
}
\label{strangefig}
\end{figure}

Unlike the charged particle density which
 scales with the number of wounded nucleons in $d+Au$
collisions, the ratio $K/\pi$ depends on the centrality. Therefore, we cannot
 assume that in the mantle the particle ratios 
are similar as in $p+p$ collisions. Preliminary data from the PHENIX
 collaboration \cite{phenixmata} show an increase of the $K/\pi$ ratio with the
mean number of binary collisions $\nu=\frac{ N_{bin}}{N_{part}/2}$.
This dependence can be fitted for the $K^-/\pi^-$ ratio by the formula
\begin{equation}
\label{scaledau}
\left(\frac{K^-}{\pi^-}\right)_{dAu}(\nu)=0.062+0.025 \times \nu \ .
\end{equation} 
Within the core-mantle model the number of produced particles can be written
as 
\begin{eqnarray}
N_K=\frac{N_{part}^{m}}{2}n_{K}^{el}+\frac{N_{part}^{c}}{2}n_{K}^{th}
\nonumber \\
N_\pi=\frac{N_{part}^{m}}{2}n_{\pi}^{el}+\frac{N_{part}^{c}}{2}n_{\pi}^{th}
\ ,
\end{eqnarray}
where $n^{el}$ and $n^{th}$ represent the production of particles per
participant pair in the mantle and in the core respectively.
Using $n_\pi^{th}\simeq (1+\alpha)n_\pi^{el}$, i.e. the pion 
multiplicity follows the charged particles scaling, we have
\begin{equation}
\frac{N_K}{N_\pi}\simeq \frac{N_{part}^{m}\frac{n_K^{el}}{n_\pi^{el}}+
(1+\alpha)N_{part}^{c}\frac{n_K^{th}}{n_\pi^{th}}}
{N_{part}^{m}+(1+\alpha)N_{part}^{c}} \ .
\end{equation}
The $K^-/\pi^-$ ratio in the thermal core should reproduce the experimental
data for the most central collisions.
We take  $\frac{n_{K^-}^{th}}{n_{\pi^-}^{th}}=0.16$ from a statistical
  hadronization model \cite{share} with parameters fitted to particle ratios 
in central collisions. The ratio
 in the mantle  $\frac{n_K^{el}}{n_\pi^{el}}$ is taken according to
the formula (\ref{scaledau}), where $\nu$ is the average number of collisions 
in the mantle part of the interaction region of a $Au+Au$ collision.
The results reproduce well the experimental data \cite{phenixstrange}
 on $K^-/\pi^-$ ratio
as a function of centrality (Figure \ref{strangefig}).
 The increase of  strange particle production
comes from the  increase of the number of collisions in the mantle and from the
increase of the contribution of the thermal core to the total emission.
Neglecting the dependence of the ratio $\frac{n_K^{el}}{n_\pi^{el}}$
or $\left(\frac{K^-}{\pi^-}\right)_{dAu}$
on the number of collisions we cannot reproduce the centrality dependence 
in $d+Au$ and $Au+Au$ interactions (dashed line in Figure \ref{strangefig}).
We note that the smooth scaling of the production  from peripheral $d+Au$ up to
central $Au+Au$ collisions  (solid line
in figure \ref{strangefig}) cannot be extended to the data for $K^+/\pi^+$.

\section{Jet absorption}

Jet quenching  has been advocated as a tool for the tomography of the dense
phase in the collision \cite{jet2}. Experimental data show a significant
suppression   of  jets in $Au+Au$ collisions but not in $d+Au$ interactions,
suggesting that the reduction of the jet rate is due to the dense medium
created in heavy-ion collisions \cite{phobosjet,phenixjet,starjet,brahmsjet}.
These results were   supported by the observation of the disappearance of back
to back azimuthal jet 
correlations in central $Au+Au$ collisions \cite{starjet}.
The origin of the jet suppression is the energy loss of jets or the absorption
of jets
 in the dense matter. In the following  we use a model of  jet
absorption \cite{jetabs} which can reproduce the jet suppression in the high
$p_\perp$ range.  

\begin{figure}
\centering
\includegraphics*[width=0.7\textwidth]{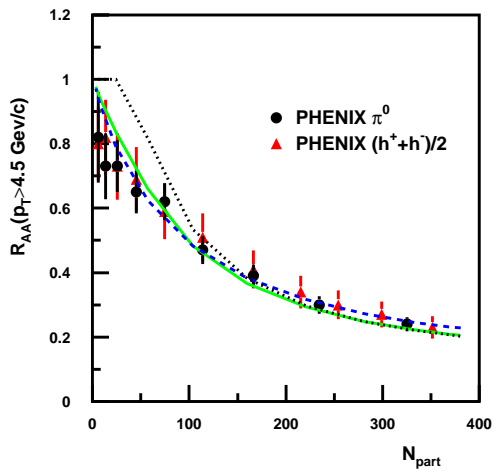}
\caption{$R_{AA}$ for charged hadrons and for $\pi^0$ integrated for
$p_\perp>4.5$GeV 
as a function of centrality 
\cite{phenixpt}. The absorption only in the thermalized source is
denoted by the dotted line. Absorption proportional to the 
 wounded nucleon density
is represented by the dashed line, the same is  for  the solid line but with an
absorption increased by $65\%$ in the thermal core.}
\label{raafig}
\end{figure}

The  survival probability
\begin{equation}
\label{abspro}
f=e^{-kI}
\end{equation} of a jet originating from
a binary collision at $(x,y)$ and traveling in the direction $(n_x,n_y)$
 depends on the density of the matter along its
trajectory 
\begin{equation}
\label{absint}
I=\int_\tau^\infty dl \rho(x+l n_x, y+l n_y) \ ,
\end{equation}
modifications due to the longitudinal expansion and the increase 
of the absorption with $l$ largely cancel out and are not taken into account.
 Taking for the density the wounded nucleon density
$\rho=d^2N_{part}/ds^2$, the jet suppression
at high transverse momentum can be reproduced \cite{jetabs}.
Using for  the formation length  of the jet   $\tau=0.2$fm 
and for the absorption
coefficient $k=0.12$fm,  the PHENIX data on jet suppression 
\cite{phenixpt} can be reproduced (dashed line in figure \ref{raafig}).

In a different scenario, it can be assumed that the absorption occurs mainly
within the dense thermal core of the collision. To illustrate the effect
we use  the  extreme 
 limit with a large  absorption within the deconfined thermal core 
and no absorption in the mantle. This corresponds to changing $\rho$ to
$\theta(d^2N_{part}/ds^2-d_{cut})$ 
in equation (\ref{absint}) and  taking
$k=0.8$fm$^{-1}$ in equation (\ref{abspro}). We use a larger value of
 $\tau=0.8$fm, which
 corresponds in  this scenario to the formation time of the thermalized
opaque core. The jet suppression in the most central collisions
is reproduced (dotted line in figure \ref{raafig}). 
In  peripheral collisions the thermal
core disappears and the jet suppression ratio $R_{AA}$ goes to one, unlike
in  the
data. If we believe that the dense phase is not 
created in peripheral collisions, this
means that the jet suppression must occur in the lower density mantle part of
the interaction region as well. This observation confirms the theoretical
calculation of the jet  suppression which increases smoothly with the density,
irrespectively of the nature of the absorbing medium \cite{baier}.

Finally we perform a calculation taking into account the increased density
within the thermalized core. Following 
the increase of the particle production we
take 
\begin{equation}
\rho=\frac{d^2N_{part}}{ds^2}\theta
\left(d_{cut}-\frac{d^2N_{part}}{ds^2}\right)+(1+\alpha)
\frac{d^2N_{part}}{ds^2}\theta
\left(\frac{d^2N_{part}}{ds^2}-d_{cut}\right)
\end{equation}
Using the absorption coefficient $k=0.09$fm ($\tau=0.2$fm)
 we reproduce the data in the
whole centrality range. This result shows that the existence of a 
 dense core within the
interaction region is compatible with the data on the jet suppression, 
if the absorption is proportional to the (increased) density in the core.

\section{Conclusions}

We study a model based on the separation of the interaction region
in $Au+Au$ collisions  in two
distinct parts~: a thermalized core in the most dense region and an
outer mantle which is not thermalized. The inner core is rapidly
thermalized when passing the region of the critical temperature and above. 
This includes both the kinetic and chemical equilibration. The 
decay of  the thermal core leads to particle production 
per participant pair larger by 65\% than in $p+p$ interactions 
at $\sqrt{s}=200$GeV. The charged particle production in the mantle
can be taken as  in a $p+p$ collision scaled by the number of participant
pairs.
The thermalized core appears at around $N_{part}=40$, we expect
modifications of the density of produced particle and increased fluctuations
in this range of centralities due to fluctuations in the size
of the thermal source; these  are not taken into account in this study.
$K^-/\pi^-$ ratio as a function of centrality can be reproduced as a sum of a
contribution of a chemically equilibrated  thermal source and 
of the production from the mantle. In the mantle an increased strangeness
per participant must be taken into account, following the $d+Au$ data. 
High $p_\perp$ jet suppression is studied in a jet absorption model. 
In order to reproduce the results in peripheral collisions, absorption must
occur in the dense core as well as in the mantle. The increase of the 
density in the thermal core resulting in an increased absorption of jets
is compatible with the data. Applying the model to $Cu+Cu$ collisions at 
$\sqrt{s}=200$GeV we find that the charged particle production 
depends on the number of participants
 similarly as in $Au+Au$ collisions, up to $N_{part}=120$. However, we expect
 that in this case the physics could be modified by fluctuations of the
 relative size of the core and of the mantle.

\bibliography{../rhic}

\end{document}